\documentclass[letterpaper, 10 pt, conference]{IEEEtran}

\usepackage{tabularx}

\usepackage{algpseudocode}
\usepackage{algorithm}
\usepackage{algorithmicx}

\usepackage{lipsum} % For algorithm smape after or before:
\usepackage{arydshln}
\usepackage[dvips]{color}
\usepackage{comment}
\usepackage{todonotes}
\usepackage{epsf}
\usepackage{epsfig}
\usepackage{times}
\usepackage{epsfig}
\usepackage{graphicx}
\usepackage{bbold}
\usepackage{mathtools}
\usepackage{mathrsfs}
\usepackage{amssymb}
\usepackage{pdfpages}
\usepackage{epstopdf}
\newfloat{algorithm}{t}{lop}

\usepackage{amsmath}
\usepackage{dsfont}
\usepackage{lettrine} % \lettrine[findent=1pt]{{{R}}}{}

\usepackage{amsmath,epsfig,amssymb,algorithm,algpseudocode,amsthm,cite,url}
\usepackage{subcaption}
\allowdisplaybreaks
\usepackage{csquotes}
\usepackage{verbatim}
\usepackage[english]{babel}
\usepackage{amsmath,amssymb}

\usepackage[justification=centering]{caption}
\usepackage{verbatim}

\begin{document}
	\title{Deep Learning for Reliable Mobile Edge Analytics in Intelligent Transportation Systems}
	\IEEEoverridecommandlockouts
	\author{\IEEEauthorblockN{Aidin Ferdowsi\IEEEauthorrefmark{1}, Ursula Challita\IEEEauthorrefmark{2}, and Walid Saad\IEEEauthorrefmark{1}\\}
		\IEEEauthorblockA{\IEEEauthorrefmark{1}
			Wireless@VT, Bradley Department of Electrical and Computer Engineering, \\ Virginia Tech, Blacksburg, VA, USA,
			Emails: \{aidin,walids\}@vt.edu\\}
			\IEEEauthorblockA{\IEEEauthorrefmark{2}
			School of Informatics, The University of Edinburgh, Edinburgh, UK., Email: {ursula.challita@ed.ac.uk}\\}\vspace{-1cm}
		\thanks{This research was supported by the U.S. National Science Foundation under Grants OAC-1541105 and IIS-1633363.}
	}
	\maketitle


	% \fi
	\IEEEpeerreviewmaketitle
	
\begin{abstract}
	Intelligent transportation systems (ITSs) will be a major component of tomorrow's smart cities. However, realizing the true potential of ITSs requires ultra-low latency and reliable data analytics solutions that can combine, in real-time, a heterogeneous mix of data stemming from the ITS network and its environment. Such data analytics capabilities cannot be provided by conventional
	cloud-centric data processing techniques whose communication and computing latency can be high. Instead, edge-centric solutions that are tailored to the unique ITS environment must be developed. In this paper, an edge analytics architecture for ITSs is introduced in which data is processed at the vehicle or roadside smart sensor level in order to overcome the ITS latency and reliability challenges. With a higher capability of passengers' mobile devices and intra-vehicle processors, such a distributed edge computing architecture can leverage deep learning techniques for reliable mobile sensing in ITSs. In this context, the ITS mobile edge analytics challenges pertaining to heterogeneous data, autonomous control, vehicular platoon control, and cyber-physical security are investigated. Then, different deep learning solutions for such challenges are proposed. The proposed deep learning solutions will enable ITS edge analytics by endowing the ITS devices with powerful computer vision and signal processing functions. Preliminary results show that the proposed edge analytics architecture, coupled with the power of deep learning algorithms, can provide a reliable, secure, and truly smart transportation environment.
\end{abstract}
\section{Introduction}\label{Intro}
Intelligent transportation systems (ITSs) are arguably the most anticipated smart city services \cite{2011Zhang}. ITSs are built on the premise of endowing vehicles and transportation infrastructure with connectivity, sensing, and autonomy, so as to provide safer road travel and ensure effective transportation. To enable this ITS vision, there is a need to equip vehicles and transportation infrastructure with smart sensors that can collect and process a heterogeneous set of data on each vehicle, its passengers, and its environment \cite{2011Zhang}. This information collection must be done at ultra-low latency and in real time, since ITSs must support autonomous features, such as self-driving vehicles. Indeed, monitoring and managing the operation of an ITS requires a reliable communication and data analytics infrastructure to transmit and process the various smart sensor data \cite{2009Papadimitratos}. However, due to the massive number of mobile sensors in an ITS, the transmission of all of the sensor measurements to a remotely located cloud -- as is done traditionally -- can result in high delays, communication network congestions, and computational overload. This will not be tolerable for an ITS since any delayed decision in the system might cause congestions, travel delays, and accidents. Therefore, efficient \emph{edge computing and analytics} frameworks must be proposed to process much of the data at the level of each individual vehicle (and infrastructure) and send only the results to the cloud in order to meet the ITS latency and reliability challenges.

Current improvements in processing units of mobile devices, that can handle Gigabits of data in real-time, make them suitable for implementation of ITS edge analytics solutions. The passengers' mobile devices can be used as edge processors due to the available power supply from the vehicle. Moreover, every vehicle can be equipped with capable processing devices such as microprocessors, that can support complex, edge computing processes. However, for effective data processing, computationally capable edge devices cannot enable the vision of ITSs, unless they are coupled with artificially-intelligent algorithms that can perform optimized edge analytics. In order to enable such smart, edge analytics, one can rely on the emerging frameworks from \emph{deep learning} \cite{chen2017machine}, that have shown their effectiveness in dealing with large image and signal datasets. Deep learning techniques can perform sophisticated functions which cannot be analytically derived such as image and speech recognition \cite{chen2017machine,speech2013Graves,Lecun2015}. Therefore, if properly deployed, deep learning tools can be an effective edge analytics tool for ITSs, as they enable the optimization and processing of the heterogeneous ITS data. 

Deep learning approaches have already been proposed for various ITS applications such as in \cite{2015Lv} and \cite{2014Huang}. In \cite{2015Lv}, the authors proposed a novel deep-learning based traffic flow prediction method based on an autoencoder model that represents traffic flow features for prediction. The work in \cite{2014Huang} evaluated the temporal and spatial patterns of transportation network flow using a multi task learning architecture. These early works have shown that deep learning algorithms outperform conventional regression and time series methods in transportation systems, thus motivating further research in this area.

\begin{figure}[t]
	\centering
	\includegraphics[width=\columnwidth]{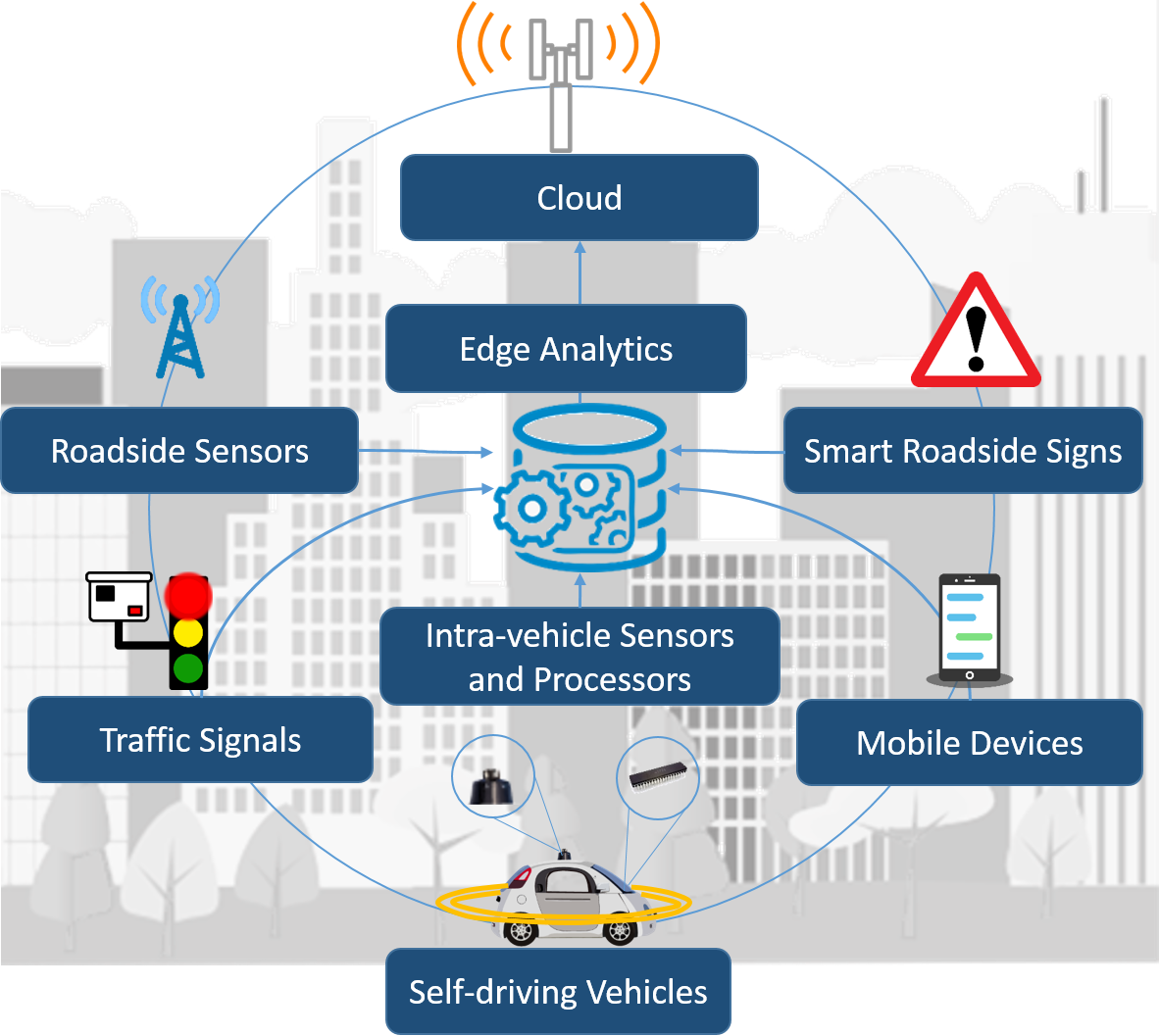}
	\caption{Proposed ITS edge analytics architecture and components.}
	\label{fig:Edge}
\end{figure}

The main contribution of this paper is to investigate the challenges and opportunities brought forward by the use of mobile edge analytics in ITSs. In particular, we first overview the data processing challenges in ITSs with a focus on latency, analytics, and computational aspects. To address these challenges, we introduce an edge computing and analytics architecture, shown in Fig. \ref{fig:Edge}, that incorporates deep learning techniques at the level of the passengers' mobile devices and intra-vehicle processors. We then propose deep learning techniques for heterogeneous data processing, path planning, vehicular platoon control, driver behavior prediction, and ITS cyber-physical security that, when implemented at the edge of an ITS, will reduce the communication and computation load on the cloud.

Here, we note that, although some surveys on deep learning and ITSs already exist such as in \cite{2011Zhang} and \cite{chen2017machine}, however, these existing surveys do not propose any ITS edge analytics framework. Moreover, these surveys do not study the reliability and latency challenges pertaining to ITS data analytics. Instead, we propose a novel ITS edge analytics architecture that can exploit deep learning techniques running at the level of passengers mobile devices and intra-vehicle processors to process large datasets and enable a truly smart transportation system operation. This architecture improves the performance of ITS in terms of reliability and latency, as it allows the transmission of only a summary or basic results of the data to the cloud, instead of transmitting the entire generated data.

The rest of this paper is organized as follows. In Section \ref{sect:ITS} we overview the ITS edge analytics challenges. Section \ref{sect:DL} proposes deep learning techniques for ITS data processing. Finally, conclusions are drawn in \ref{sect:conc}.
\section{Mobile Edge Analytics in ITSs: Opportunities and Challenges}\label{sect:ITS}
ITSs will encompass a broad range of mission-critical data analytics tasks that include autonomous control, path planning, and vehicular platooning. Self-driving vehicles will be equipped with numerous sensors that will collect measurements from each vehicle's environment. Moreover, roadside sensors and traffic control signals will become more intelligent and will generate high volumes of data. The information generated from these sensors must be processed with low latency to ensure a reliable ITS operation. To operate in an ultra-low latency regime, typically, each sensor must generate shorter packet sizes which, when combined with a massive number of sensors, will cause severe degradation in communication throughput. On the other hand, to enhance reliability, more communication resources are needed. Therefore, given the limited availability of wireless communication and computational resources, achieving the latency and reliability requirements of ITSs will be challenging. 

One conventional approach is to rely on a remote cloud to upload data and run sophisticated algorithms. However, this approach yields to high end-to-end latency due to the massive computation and communication load generated by the vehicles and roadside sensors. Therefore, the use of an edge computing architecture, such as the one proposed in Fig. \ref{fig:Edge}, that relies on highly capable intra-vehicle processors as well as passengers mobile equipment to perform ITS edge analytics can potentially improve both latency and reliability, while reducing the reliance on a remote cloud. To enable such edge analytics in the context of ITS, one must meet a number of challenges that range from heterogeneous data integration and vehicular dynamics control to driver behavior prediction. Next, we provide an in-depth exposition of those challenges and we propose a deep learning framework to address them.
\subsection{Heterogeneous Data Sources}
Developing autonomous and connected ITSs will require the use of different sensor types to collect a heterogeneous mix of measurements including environmental features, vehicle dynamics, fuel consumption, and driver's fatigue level\cite{FAOUZI20114,2011Zhang}. The heterogeneity and sheer volume of such real-time information will require high reliability and ultra-low latency for mutual information sharing between vehicles or event-driven and cyclic reports to the cloud. To improve reliability and latency requirements, optimal processing methods must be deployed to combine such heterogeneous data at the level of each vehicle to reduce the unnecessary and redundant information before transmission to other vehicles or to the cloud.
\subsection{Path Planning and Autonomous Control}
One key ITS challenge is the optimal routing of self-driving vehicles in an effort to reduce the total road travel delay\cite{FU20063324}. Moreover, self-driving vehicles must control their speed, direction, and acceleration autonomously \cite{2014Gerla}. Given the high density of the vehicles in crowded streets and the random decisions taken by human drivers, performing vehicular path planning and autonomous control algorithms will be challenging. Although a centralized cloud can update the path plan and control decision of each self-driving vehicle this requires real-time tracking of the vehicles as well as the transmission of location information from the vehicles to the cloud. Such a cloud-centric approach will not be suitable for ITS due to the need for high bandwidth to transmit raw data as well as the increased latency over the the vehicle-to-cloud communication link. To overcome these challenges, online edge algorithms must be implemented individually by each vehicle to plan the future path and control the vehicle's dynamics.
\subsection{Vehicular Platoon Control}
Platooning of self-driving or driver-controlled vehicles can potentially enhance highway safety, improve traffic stability, and reduce fuel consumption\cite{2015Li}. In a platoon, each joining self-driving vehicle must individually take decisions regarding its dynamics with the least effect on the stability of the platoon. The decisions taken by each vehicle require real-time processing and collection of measurements and images from its environment. A key point here is that, in a platoon, each vehicle experiences approximately a same environment and collects almost identical measurements, with a small delay compared to its leading vehicle. Hence, mutual information sharing between the vehicles inside a platoon can potentially reduce the total computational load in the platoon due to redundant measurements. Moreover, platoons provide an ideal use case for collaborative edge analytics such as \emph{federated learning} in which vehicles collaboratively learn a shared prediction model while keeping all of the training data on their own which helps them to be decoupled from the cloud. In a nutshell, using edge analytics to reduce a platoon's total data generation will enable ultra-low latency and reliable operation as it substantially reduces the total communication and computation overhead.
\begin{figure*}[h]
	\centering
	\includegraphics[width=\textwidth]{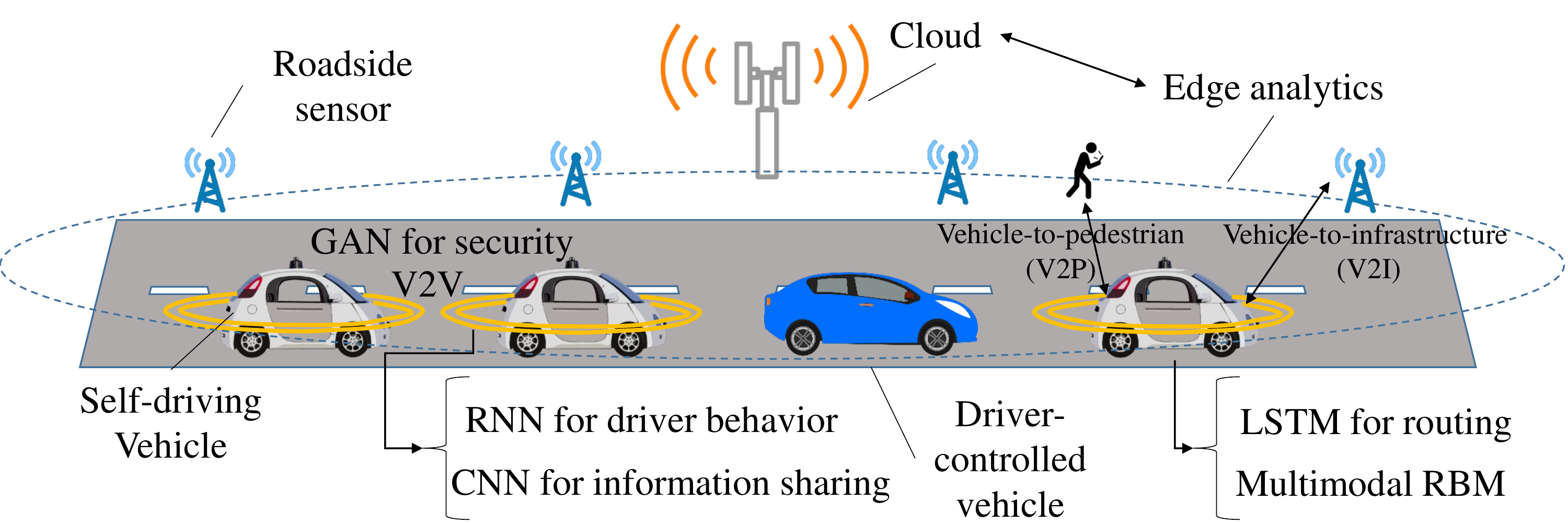}
	\caption{An ITS that incorporates the various proposed edge analytics and deep learning methods.}
	\label{fig:ITS}
\end{figure*}
\subsection{Semi-autonomous ITSs}
Next-generation \emph{semi-autonomous ITSs} must ensure the co-existence of both autonomous, self-driving vehicles, as well as driver-controlled vehicles. While autonomous vehicles can use sensing technologies to optimally navigate to their destination, driver-controlled vehicles will still rely on humans, and, hence, their decisions will often be guided by potentially unpredictable behavior\cite{Dresner2007}. This co-existence between autonomous controlled vehicles and the unpredictable behavior of human drivers will bring forward new edge analytics challenges. For instance, the existence of drivers with unpredictable actions will cause more sensor readings in the autonomous vehicles that must track the drivers' decisions and optimize their own dynamics. Thus, novel edge analytics frameworks are needed to identify self-driving vehicles from driver-controlled vehicles and predict the behavior of a driver in the vicinity of other self-driving cars. The prediction of driver behavior will reduce the computational overhead due to the reduction in the sensor readings of self-driving vehicles in the proximity of driver-controlled vehicles. Indeed, this computational load reduction will allow for low-latency edge analytics.
\subsection{ITS Security}
Along with their performance benefits, edge analytics techniques can make the ITS more vulnerable to cyber-physical attacks as adversaries will find it easier to attack the link between vehicles rather than attacking a more secure cloud. Therefore, security of information sharing across ITS components such as vehicles, roadside sensors, and the cloud is crucial since any faulty sensor measurement and process at the ITS edge can cause accidents and injuries\cite{2009Papadimitratos}. Moreover, a cyber attacker that can inject faulty information to a platoon or autonomous vehicle will cause non optimality in their control system stability. Therefore, there is a need for new techniques to authenticate and secure the communication links between vehicles, as well as between vehicles and cloud. For instance, to implement real-time edge analytics, inter-vehicle communication must be properly secured against cyber attacks in the vicinity of the vehicles. Furthermore, the ITS must be resilient to possible collisions so as to recover from congestions caused by accidents.  Hence, online vehicle dynamics tracking algorithms should be studied to detect anomalous ITS device behavior and predict possible collisions.

Managing massive ITS generated data for path planning, vehicular platoon control, and driver behavior identification requires computer vision techniques to extract image features, identify the environment, and detect the surrounding objects. Processing heterogeneous ITS data involves extraction of correlations between the signals of different sources. From a security perspective, relying on conventional message protection techniques such as encryption will not be effective for inter-vehicle communication links, since the transmitted message between vehicles will be of small size. Hence, we propose \emph{deep learning techniques} for meeting the ITS edge analytics challenges due to their proven effectiveness\cite{Lecun2015}. Deep learning allows computational methods that are composed of multiple processing layers to learn data features with higher levels of abstraction. Next, we first introduce the basics of deep learning and, then, discuss how deep learning can be used for ITS edge analytics.
\section{Deep Learning for Mobile Sensing in ITS}\label{sect:DL}
\subsection{Overview on Deep Learning}
Deep neural networks (DNNs) are an emerging type of artificial neural networks (ANNs). In general, ANNs are function approximators used to recognize correlations between relevant features and the output of processed data. % They mainly consist of an input layer that represents the input signal, at least one hidden layer that consists of a number of neurons used to map the output of the previous layer to another space that is fed to the next layer, and an output layer that consists of a logistic, or softmax, classifier that assigns a likelihood to a particular outcome.%
They consist of an input layer that represents the input signal, at least one hidden layer, and an output layer that consists of a logistic, or softmax, classifier that assigns a likelihood to a particular outcome. DNNs are ANNs with multiple hidden layers\cite{chen2017machine}. In DNNs, each hidden layer will train on a distinct features set based on the previous hidden layer's output. This is known as feature hierarchy and allows DNNs to model high-level abstractions in data thus learning multiple levels of representation and abstraction. Therefore, DNNs are suitable for recognizing and modeling complex features from large and high-dimensional datasets with many parameters such as camera, LIDAR, radar, and vehicle dynamics readings in ITSs. This renders DNNs suitable for meeting the edge analytics challenges of ITS discussed in Section II. Several types of DNNs can be be used for ITS edge analytics such as deep convolutional neural networks (CNNs) and restricted Boltzmann
machines (RBMs) that can be used in heterogeneous data processing and automatic control, recurrent neural networks (RNNs) and that can be applicable for vehicle dynamics estimation, and generative adversarial networks (GANs) and long-short term memory (LSTM) that are useful for ITS security. Next, we provide an in-depth look on how such DNN techniques can be used for reliable ITS edge analytics.

\begin{figure*}[h]
	\centering
	\includegraphics[width=\textwidth]{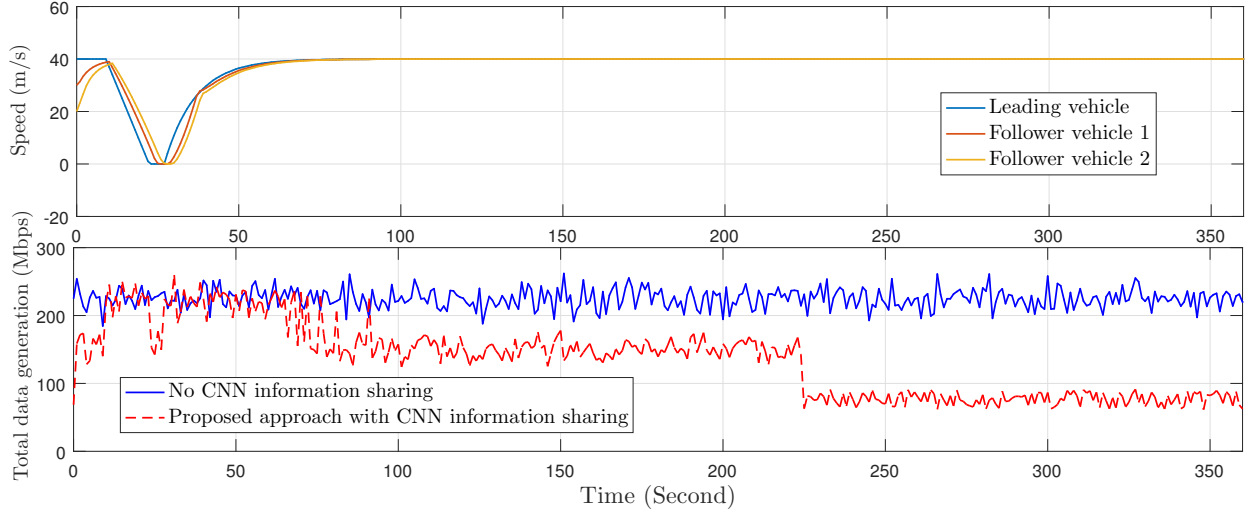}
	\caption{Computational load in an SVN where vehicles share similar information from the environment.}
	\label{fig:sim1}
	\vspace{-1mm}
\end{figure*}

\subsection{Deep Learning for ITS Edge Analytics and Mobile Sensing}

\subsubsection{Multimodal RBM for Heterogeneous Data Integration}
Self-driving cars will encompass a variety of sensors, from light sensors and cameras to ultrasound sensors, enabling each vehicle to make sense of its surrounding environment.  As such, any control action taken by a self-driving vehicle will depend on the different types of sensor data. Integrating such heterogeneous sensor readings into one vector can provide a better assessment of the environment as compared to using each type of data independently. Nevertheless, there exists differences between sensors ranging from sampling rates to the data generation model thus making ITS sensor integration challenging. In this regard, multimodal RBMs are a suitable tool for combining different perspectives captured in signals of multimodal data for a system with multiple sensors \cite{2012Srivastava}. A multimodal RBM can be implemented in the ITS edge components to identify nonintuitive features largely from cross-sensor correlations which can yield accurate estimation\cite{2011Zhang}. For instance, a system trained simultaneously to detect lane markings, cars, and pedestrians does better than three separate systems trained in isolation since the single network can share information among the separate tasks.  

\subsubsection{CNN-LSTM for Path Planning and Autonomous Control}
Autonomous self-driving cars require a perception system to sense their surrounding environment such as other cars, people, lampposts, and curbs. Therefore, to determine the paths of self-driving vehicles, one can combine online and offline solutions into one framework. Self-driving vehicles can use an offline environmental map as one of many data streams, combining it with other sensor inputs that detect dynamic objects to better determine their paths. This, in essence, allows self-driving vehicles to take actions in an online manner and at the edge thus reducing the delay for decision making and the probability of occurrence of accidents or congestion events while increasing fuel efficiency. In this regard, CNNs can be combined with LSTMs within self-driving vehicles, as a part of the introduced edge analytics architecture, resulting in a CNN-LSTM network. In particular, CNN can be used to identify objects by extracting the features from the input image \cite{Lecun2015}. These features are then fed to an LSTM network which is used in the sequence task portion of the system\cite{speech2013Graves}. This will allow self-driving cars to take a sequence of steps along their corresponding paths based on the features extracted by the CNN.

%
%https://devblogs.nvidia.com/parallelforall/deep-learning-self-driving-cars/

%https://www.technologyreview.com/s/602600/deep-driving/
%
%https://arxiv.org/pdf/1705.08624.pdf

\subsubsection{CNNs and Vehicular Communication for Platoon Control}
In a platoon, vehicles following a leading vehicle, experience an almost identical road environment as the leading vehicle. Therefore, instead of processing the collected data at all of the vehicles, only the leading vehicle can process the environment and share the result with its followers using vehicle-to-vehicle (V2V) communication links \cite{2009Papadimitratos}.

Two fundamental requirements for this procedure are: a) a mechanism for similarity detection in the collected data of different vehicles and b) a vehicular communication network for sharing the results among vehicles. A CNN can allow a vehicle to compare its own environmental observations with those received from the leading vehicle, and in case of similarity the processing result from the leading car can be used. Therefore, in summary, a leading vehicle passing a stationary roadway processes its observations at the edge and can share the result with the following vehicles and roadside infrastructure using a cellular network. To this end, cellular communication technology can support vehicle-to-everything (V2X) communications\cite{2009Papadimitratos}. The following vehicles compare the leading vehicle's observation with their own observations through a CNN and in case of similarity, the following vehicles use the result of the leading vehicle without processing their own observation. This in turn reduces the computational overhead caused by processing the observation of all the vehicles at the cloud. CNNs are also suitable for various wireless V2X resource allocation problems in ITS. In essence, many resource allocation functions, such as power control or spectrum allocation can make use of CNN-based reinforcement learning algorithms to dynamically adapt the V2X network resources to the operation of the vehicles.
\begin{figure*}[h]
	\centering
	\includegraphics[width=\textwidth]{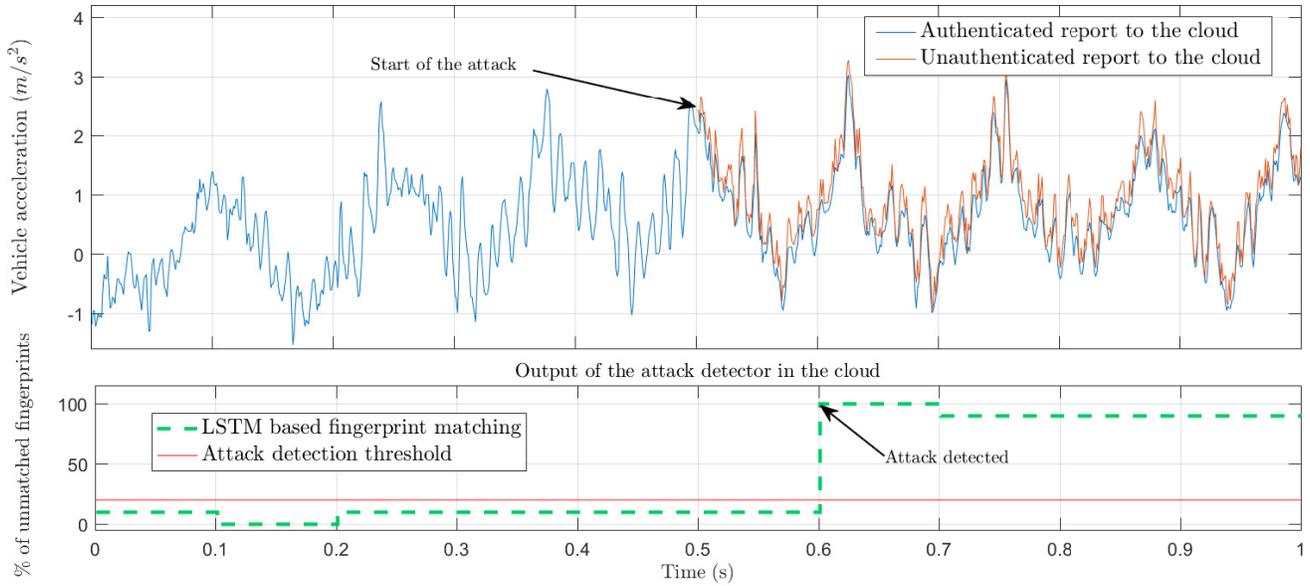}
	\caption{LSTM for cyber attack detection in the cloud.}
	\label{fig:attackdetection}
\end{figure*}
\begin{figure*}[h]
	\centering
	\includegraphics[width=0.9\textwidth]{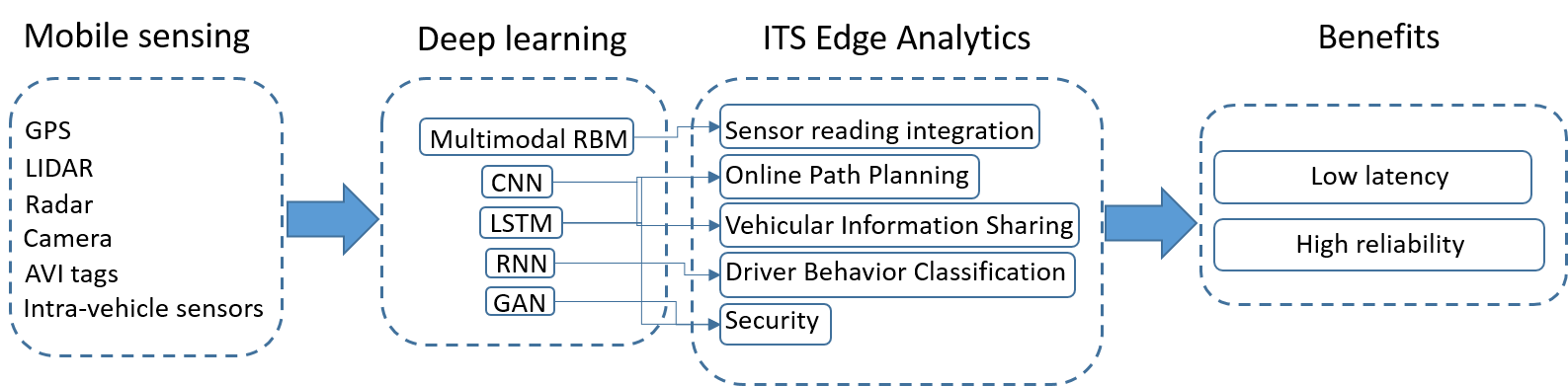}
	\caption{Deep learning architecture for mobile sensing in ITS.}
	\label{fig:DL}
	\vspace{-5mm}
\end{figure*}

To demonstrate the benefits of information sharing and CNNs, we investigated the decision sharing of a leading vehicle in a platoon of self-driving vehicles. Fig. \ref{fig:sim1} shows a simulation of three self-driving vehicles travelling over the same road. Two cases are considered: a) All vehicles independently record images from the environment and, using their own CNN, decide on their speed and wheel angle and b) Only the leading vehicle records images from the environment and takes decision using its own CNN and shares the decision with its followers. The followers use the decision shared by the leading vehicle without recording images from the environment when they are at a \emph{steady state}, however, they use their own CNN while they are not at a steady state. The steady state is a system state at which all of the platoon's vehicles have an equal speed. Before $ 75 $ seconds, all three vehicles are at a non-steady state and, thus, they all observe their environment using their own cameras resulting in $ 250 $ Mbps of generated data. However, after $ 75 $ seconds, we can see that, since leading vehicle and first following vehicle are at a steady state, the first follower uses the computational results of the leading vehicle without using its own camera. Therefore, the total data generation reduces to $ 150 $ Mbps. After $ 225 $ seconds, the second follower reaches the steady state and uses the shared results of the leading vehicles and stops using its own camera. Consequently, the total data generated by the platoon reduces to $ 75 $. Fig. \ref{fig:sim1} demonstrates that information sharing by a leading vehicle results in a reduction of total sensor readings in the platoon.

\subsubsection{RNN for Driver Behavior Prediction}
An RNN is an ANN that can process a sequence of input to extract temporal behaviors \cite{2014Sutskever}. In a semi-autonomous ITS, the driver behavior can be predicted using RNN blocks implemented at the edge, i.e., at the self-driving vehicles following a driver-controlled vehicle. Here, a following self-driving vehicle will observe the decisions taken by the leading driver in different situations and will use these observations as inputs to an RNN. Consequently, the RNN will recognize the behavior of the leading driver and will predict its future actions. Eventually, prediction of the leading driver behavior will reduce the computational load on a following self-driving vehicle since less sensor readings from the leading driver-controlled vehicle will be needed.

\subsubsection{GAN and LSTM for ITS Security}
GANs are ANNs that learn to create synthetic data similar to some real data. GANs consist of two main networks: a generative network and a discriminative network\cite{2014Goodfellow}. The discriminator is a classifier that distinguishes between artificially generated data and real data. It can take the form of a CNN binary classifier. While the discriminator network identifies the reality of a dataset, the generative network takes random input values and transforms them into stealthy datasets that are not recognizable by the discriminative network. In an ITS, a cyber attacker can transmit a stealthy message to self-driving vehicles and manipulate their decisions. In this respect, a GAN's discriminator network can be implemented at the ITS edge components to distinguish fake data generated by an adversary and real data.

Moreover, LSTM is a suitable ANN for data transmission security since it can extract information from a sequence of past data and predict the future data sequence\cite{2014Sutskever}. LSTMs can be trained to fingerprint a data sequence which allows authentication of messages between vehicles. In the case of unauthenticated messages from a cyber attacker, fingerprints extracted from the data sequence will not match with authenticated fingerprints. Fig. \ref{fig:attackdetection} shows a comparison between authenticated and unauthenticated reports transmitted from a vehicle to the cloud. Here, one LSTM network is implemented to extract the signal fingerprint and then embed it in the signal. At the cloud, another LSTM network is implemented to extract fingerprints from the reported data and compare them with the embedded authenticated fingerprint set. Based on the difference between the extracted and authenticated fingerprints, cyber attacks can be detected. Fig. \ref{fig:attackdetection} shows that the security check at the cloud causes a delay of only $ 0.1 $ seconds which can be reasonable for an ITS. GAN and LSTM networks can solve reliability and latency challenges in ITS security applications by being implemented at the edge of ITS since they will have better accuracy with lower delay compared to traditional cloud-based security architecture. 

Fig. \ref{fig:DL} briefly overviews the various proposed deep learning solutions.
\subsection{Future Research Challenges in DNNs for ITS Edge Analytics}
DNNs are a promising solution for addressing a wealth of ITS edge analytics and sensing problems. However, there exists a number of challenges that must be overcome in future research, as outlined next:
\subsubsection{Data Accuracy} Deep learning techniques require a large set of accurate data for training. Inaccurate datasets used for training will result in an erroneous ANN. Since the decision making process in an ITS requires heterogeneous types of data including vehicle's position and speed, spacing between the vehicles and distance to the roadside objects, therefor data collection needs a precise supervision. Moreover, since in the edge architecture, vehicles, roadside sensors, and mobile devices process the data, results can be more error prone due to the limited capabilities of edge devices.
\subsubsection{Unpredictable Situations}An ITS might encounter unpredictable situations due to the existence of randomness in drivers' behavior and environmental effects. Such unpredictable situations will cause reaction delay in the system. Therefore, ITS-centric deep learning methods must account for unpredictable situations by learning the behavior of drivers to reduce the delay caused by reacting to drivers' action. In this case, since the driving profiles cannot be stored on the edge of the ITS due to the security and memory constraints, the number of queries to the cloud will increase.
\subsubsection{Resilience}ITS DNNs must be resilient to vehicle accidents. An ITS must recover from accidents as quickly as possible to avoid long transportation delays. This is particularly important when processing is done at the edge, as proposed here, since edge devices may be damaged and, thus, they need to rely on either neighbors or revert back to cloud processing. Therefore, possible accidents must be taken into account while implementing deep learning methods at the edge of the ITS.
\subsubsection{Drivers against Self-driving Vehicles}Although an RNN is useful in driver behavior classification and prediction, driver-controlled vehicles might also change their behavior while interacting with self-driving vehicles. To address this issue, game-theoretic solutions can be developed in which the players are self-driving vehicle and a driver-controlled vehicle sharing a road. The optimal distance and speed for these vehicles can be derived by taking into account their dependent actions while they are building a platoon. A key advantage of game-theoretic approaches is their inherent ability to be implemented in a self-organizing manner, at the ITS edge. Therefore, by combination of game-theory with an RNN, self-driving and driver-controlled vehicles interaction can be analyzed at the ITS edge.
\subsubsection{Robustness}DNNs can be compromised by intentional noise injected by cyber attackers. For instance, altering some of pixels in a traffic sign might deceive an ANN from acting properly. In general, moving the intelligence to the network's edge will increase the threats of cyber attacks, and, hence, any proposed deep learning methods must be properly secured. 
\subsubsection{Delay-intensive Training Phase}The training phase of a DNN can be time consuming which makes their usage and testing nontrivial in some scenarios, especially, when trying to add a new feature to the DNN. Hence, better training methods must be proposed to reduce the training delay. For example, federated learning approaches where the edge components can cooperatively learn the features can boost the train performance.
\section{Conclusion}\label{sect:conc}
In this paper, we have proposed a novel ITS architecture that relies on edge analytics and deep learning to optimize its computation, latency, reliability, and overall operation. In particular, we have proposed deep learning techniques to solve a plethora of edge analytics challenges in ITSs that include measurements heterogeneity, path planning, autonomous vehicle control, platoon control, semi-autonomous ITSs, and cyber-physical security. For each challenge, we have proposed a suitable deep learning framework that can be implemented as a part of an edge computing and analytics architecture to process a vehicle's data before transmitting to the cloud. Finally, we have outlined some of the key open problems regarding the implementation of deep learning techniques for edge analytics in ITSs.
\bibliographystyle{IEEEtran}
\bibliography{references}
	
	%\vfill
	


\end{document}